\documentclass[reprint,aip,twocolumn,cha,showkeys,showpacs,superscriptaddress]{revtex4-1}
\usepackage{amsmath}
\usepackage{graphicx}% Include figure files
\usepackage{dcolumn}% Align table columns on decimal point
\usepackage{bm}% bold math
\usepackage{hyperref}% add hypertext capabilities
\usepackage{epstopdf}
\usepackage{color} % farbiger Text
\usepackage{SIunits}
\usepackage{ulem} % durchstreichen

\begin{document}

\title{Determination of effective mechanical properties of a double-layer beam by means of a nano-electromechanical transducer}

\author{Fredrik Hocke}
\affiliation{Walther-Mei{\ss}ner-Institut, D-85748 Garching, Germany}
\affiliation{Nanosystems Initiative Munich, D-80799 M\"{u}nchen, Germany}
\affiliation{Technische Universit\"{a}t M\"{u}nchen, D-85748 Garching, Germany}

\author{Matthias Pernpeintner}
\affiliation{Walther-Mei{\ss}ner-Institut, D-85748 Garching, Germany}
\affiliation{Nanosystems Initiative Munich, D-80799 M\"{u}nchen, Germany}
\affiliation{Technische Universit\"{a}t M\"{u}nchen, D-85748 Garching, Germany}

\author{Xiaoqing Zhou}
\affiliation{\'{E}cole Polytechnique F\'{e}d\'{e}rale de Lausanne, CH-1015 Lausanne, Switzerland}

\author{Albert Schliesser}
\affiliation{\'{E}cole Polytechnique F\'{e}d\'{e}rale de Lausanne, CH-1015 Lausanne, Switzerland}
\affiliation{Niels Bohr Institute, University of Copenhagen, DK-2100 Copenhagen, Denmark}

\author{Tobias J. Kippenberg}
\affiliation{\'{E}cole Polytechnique F\'{e}d\'{e}rale de Lausanne, CH-1015 Lausanne, Switzerland}

\author{Hans Huebl}
\email[]{hans.huebl@wmi.badw.de}
\affiliation{Walther-Mei{\ss}ner-Institut, D-85748 Garching, Germany}
\affiliation{Nanosystems Initiative Munich, D-80799 M\"{u}nchen, Germany}

\author{Rudolf Gross}
\email[]{Rudolf.Gross@wmi.badw.de}
\affiliation{Walther-Mei{\ss}ner-Institut, D-85748 Garching, Germany}
\affiliation{Nanosystems Initiative Munich, D-80799 M\"{u}nchen, Germany}
\affiliation{Technische Universit\"{a}t M\"{u}nchen, D-85748 Garching, Germany}

\date{\today}

\begin{abstract}
We investigate the mechanical properties of a doubly-clamped, double-layer nanobeam embedded into an electromechanical system. The nanobeam consists of a highly pre-stressed silicon nitride and a superconducting niobium layer. By measuring the mechanical displacement spectral density both in the linear and the nonlinear Duffing regime, we determine the pre-stress and the effective Young's modulus of the nanobeam. An analytical double-layer model quantitatively corroborates the measured values. This suggests that this model can be used to design mechanical multilayer systems for electro- and optomechanical devices, including materials controllable by external parameters such as piezoelectric, magnetrostrictive, or in more general multiferroic materials.
\end{abstract}

 %Microwave circuits


\maketitle

Over the last decade there has been growing interest in micro- and nanomechanical systems taylored for fundamental quantum experiments and sensing applications~\cite{Masmanidis2005, Feng:2008a, Zolfagharkhani2008, Naik:2009a, Chaste2011, Hanay:2012a}. For sensing applications, an effective signal transduction from the mechanical to the electrical domain is highly desirable to allow their integration into more complex structures. To this end, a powerful strategy for establishing extremely sensitive detection schemes is to couple the mechanical resonator to an electromagnetic circuit, leading the field of circuit-electromechanics, a subfield of optomechanics~\cite{Aspelmeyer2013,aspelmeyer_cavity_2014}. For example, an effective coupling is realized by making the mechanical entity such as a nanobeam or a micromembrane part of an on-chip microwave cavity~\cite{Regal2008,Hertzberg2009a,Massel2011,Teufel2011}. With such electromechanical hybrid systems, ground state cooling~\cite{Teufel2011a}, electromechanically induced transparency and absorption~\cite{Teufel2011,Hocke2012}, generation of slow light~\cite{Zhou2013}, state transfer of (classical) photonic states to the mechanical mode~\cite{Palomaki2012} as well as parametric amplification~\cite{Massel2011} have been demonstrated. In addition, electromechanical systems driven by a microwave  tone resonant with the cavity provide a powerful toolbox for studying nanomechanical oscillators~\cite{Sulkko2010, Hoch2011}. A recent example is the characterization of the coupling between in- and out-of-plane modes in a nanomechanical beam~\cite{Faust2012a}. Whereas for most applications only the linear regime of mechanical oscillators is relevant, the nonlinear regime is of particular interest for the study of their fundamental mechanical properties. Therefore, the Duffing  nonlinearity~\cite{Sattler2011} has been explored extensively in various nanomechanical systems~\cite{Kozinsky2006,Kozinsky2007,Karabalin2009a, Unterreithmeier2009a,Venkatesan2010,Juillard2010,Jun2010,Unterreithmeier2010b}.

In this letter, we present a systematic study of a nano-electromechanical system consisting of a doubly clamped Si$_3$N$_4$/Nb bilayer nanobeam embedded into a Nb superconducting coplanar waveguide (CPW) microwave cavity. We perform sideband spectroscopy of a probe field centered around the resonant frequency ($\sim 6$\,GHz) of the microwave cavity and use an additional AC drive tone centered around the mechanical resonance frequency ($\sim 1.5$\,MHz) to drive the mechanical amplitude in the Duffing regime. Hereby, we obtain the mechanical eigenfrequency and the ''backbone'' curve, which allow us to deduce the effective mechanical stress $\sigma_{\mathrm{eff}}$ and Young's modulus $E_{\mathrm{eff}}$. Since the nanobeam consists of a Si$_3$N$_4$/Nb bilayer, we obtain only effective mechanical parameters. We show that the mechanical behavior of a pre-stressed nanobeam consisting of two or more layers can be well described by effective material constants. Moreover, we derive how these effective parameters are related to the material constants of the individual layers. The effective material constants determined in our experiments quantitatively agree well with those obtained from model calculations.

The eigenfrequency of a doubly-clamped mechanical beam assuming hinged ends and neglecting effects due to bending at the clamping points is given by~\cite{Timoshenko2008,Verbridge2006}
\begin{equation}
\Omega_{\rm{m}} = \frac{\pi^2}{L^2} \sqrt{\frac{EI}{\rho A}\left(1+\frac{\sigma A L^2}{EI\pi^2}\right)} \approx \frac{\pi}{L} \sqrt{\frac{\sigma}{\rho}} \; .
\label{equ:Sample_Om2}
\end{equation}
Here, $\sigma$ is the stress present in the beam, $\rho$ its mass density, $E$ its Young's modulus and $I$ its moment of inertia. The the right-hand side of (\ref{equ:Sample_Om2}) approximates $\Omega_{\rm{m}}$ for high stress. In this case (\ref{equ:Sample_Om2}) allows to directly determine the stress of the beam for known $\rho$. For the more common case of a pre-stressed, doubly-clamped beam, including the bending present at the clamping points no analytical expression to derive $\Omega_{\rm{m}}$ exists (see supplemental material \cite{SI}). Nevertheless, usually  (\ref{equ:Sample_Om2}) is a good approximation, since the deviation from the exact numerical solution is typically small.

In the linear (harmonic) regime, the oscillation frequency is independent of the oscillation amplitude. In contrast, for large amplitude excitation nonlinear effects become relevant. We will show that these nonlinear effects allow us to determine the effective Young's Modulus. To account for nonlinearities, the harmonic potential has to be extended by terms going beyond the quadratic term in the displacement $x$. Due to the symmetry of the doubly-clamped nanobeam all odd parity terms can be omitted.  The next relevant term is proportional to $x^4$, leading to the additional restoring force term $\alpha x^3$ in the equation of motion. Here,  $\alpha$ is the so-called Duffing parameter characterizing the nonlinear dependence between force and mechanical displacement. The full equation of motion of a damped Duffing oscillator with linewidth $\Gamma_{\rm m}$ thus reads
\begin{equation}
     \ddot{x}(t)+\Gamma_{\rm m}\dot{x}(t)+\Omega_{\rm m}^2 x(t)+\frac{\alpha}{m_{\rm{eff}}} x^3(t) = \frac{F}{m_{\rm{eff}}} \cos(\omega t)\; .
     \label{equ:Duffing}
\end{equation}
Here, $m_{\rm{eff}}$ is effective mass of the nanobeam and the term on the right-hand side represents a harmonic driving force with frequency $\omega$ and amplitude $F$.

\begin{figure}[tb]%
\center{\includegraphics[width=0.8\columnwidth]{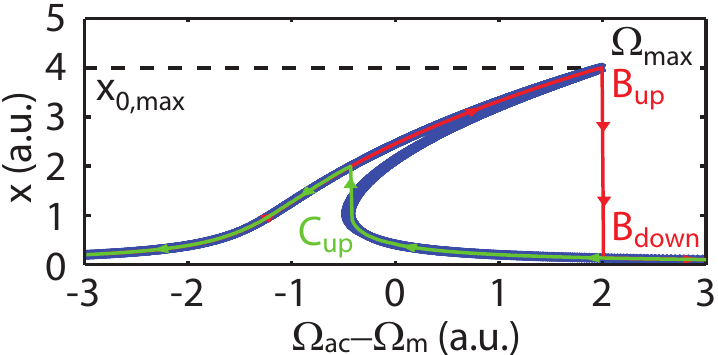}}
 \caption{
 A simulated amplitude spectrum of a Duffing oscillator with an amplitude beyond the critical one showing  hysteretic behaviour. The red (green) line indicates a frequency up (down) sweep. The blue line between the points $B_{\mathrm{up}}$ and $C_{\mathrm{up}}$ shows the metastable branch which not experimentally accessible. }
\label{Hocke_Figure1}
\end{figure}

Figure~\ref{Hocke_Figure1} shows a generic amplitude spectrum of a Duffing oscillator (blue line) obtained from (\ref{equ:Duffing}) for $\alpha>0$ (which is the common case in doubly clamped nanobeams~\cite{NonlinearBook2}) and a drive sufficiently strong to generate amplitudes exceeding the critical amplitude $x_{\rm{crit}}$. In this non-linear regime, the amplitude shows a hysteretic behavior between $C_{\rm up}$ and $B_{\rm up}$. Here, three amplitude solutions to (\ref{equ:Duffing}) exist with only two of them being stable \cite{NonlinearBook}. The red curve Fig.~\ref{Hocke_Figure1} shows the amplitude spectrum for increasing drive frequency. At the frequency $\Omega_{\rm max}(x_{\rm 0,max})$, the end point $B_{\rm up}$ of this upper branch of the Duffing oscillator, the amplitude takes its maximum value $x_{0,\mathrm{max}}$. On further increasing the drive frequency, the amplitude drops discontinuously to the low excitation state $B_{\rm down}$. The lower branch (green curve) can then be observed for decreasing drive frequencies. Here, the position of the discontinuity is at $C_{\rm up}$. Note that point  $B_{\rm up}$ can only be accessed on increasing the drive frequency.

Although eq.~(\ref{equ:Duffing}) cannot be solved analytically, one can show that $\Omega_{\rm max}$ shifts with $x_{\rm 0,max}$ according to~\cite{NonlinearBook2}
\begin{equation}
    \Omega_{\rm{max}} = \Omega_{\rm m} + \frac{3\alpha}{8m_{\rm{eff}}\Omega_{\rm m}} \; x_{\rm{0,max}}^2 \; .
    \label{equ:DuffingBackbone}
\end{equation}
This relation is called the backbone of the Duffing oscillator. It allows to extract $\alpha$ by measuring $\Omega_{\rm max}$ as a function of $x_{\rm 0,max}$. Experimentally, the difficult part is to precisely determine $x_{\rm 0,max}$ in a calibrated measurement. Having determined $\alpha$, we can use \cite{Unterreithmeier2009a}
\begin{equation}
\alpha = m_{\rm{eff}}\pi^4 \; \frac{E+\frac{3}{2}\sigma}{4L^4\rho} \;
\label{equ:Duffing_physical2}
\end{equation}
to relate the Duffing parameter to the mechanical properties of the nanobeam. In particular, we can estimate the effective Young's modulus of the nanobeam, since the stress $\sigma$ can be determined independently from (\ref{equ:Sample_Om2}) by measuring the  eigenfrequency $\Omega_{\rm m}$ in the linear regime. Additionally, we can extract the critical amplitude $x_{\rm{crit}}$, which is defined as the oscillation amplitude for coinciding $C_{\rm up}$ and $B_{\rm down}$, given by
\begin{equation}
    x_{\rm{crit}} = \left(\frac{4}{3}\right)^{3/4} \; \sqrt{\frac{m_{\rm{eff}}\Gamma_{\rm m}\Omega_{\rm m}}{\alpha}} \; .
    \label{equ:DuffingParam}
\end{equation}

\begin{figure}[tb]%
\center{\includegraphics[width=0.7\columnwidth]{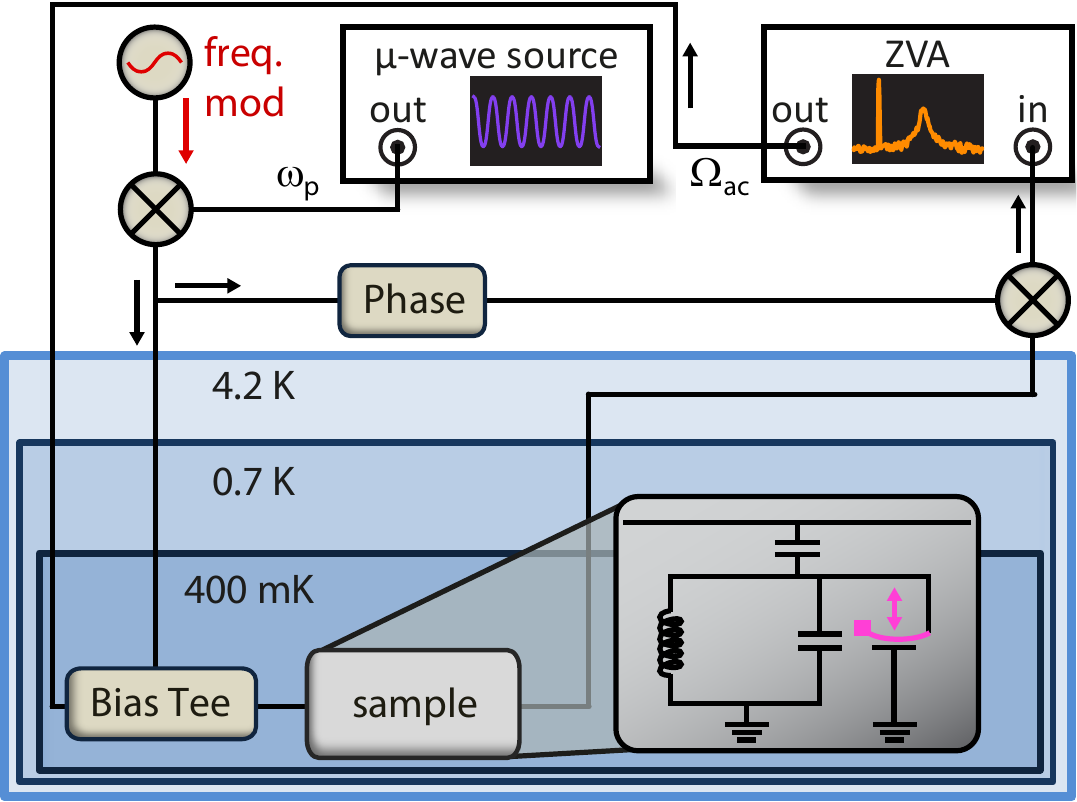}}
 \caption{Experimental setup used to determine the Duffing parameter. A vector network analyzer (ZVA) drives the nanobeam around its resonance frequency $\Omega_{\mathrm{m}}$ and records its motion by employing a homodyne detection of the transmitted continuous microwave pump tone, generated by a microwave source. The frequency of the microwave pump tone equals the $\mu$-wave cavity eigenfrequency to avoid any disturbance of the mechanical motion. The inset shows a circuit diagram of the sample, including the $\mu$-wave cavity and the doubly-clamped nanobeam capacitively coupled to it.}
\label{Hocke_Figure2}
\end{figure}

Before discussing the experimental data, we first provide basic information on the sample fabrication and the experimental setup. For the fabrication of the nano-electromechanical hybrid, we globally remove the $t_{\rm{SiN}}=70\,\nano\meter$ thick highly stressed silicon nitride layer deposited on a silicon substrate except for the area of the nano-string. Subsequently, we deposit a $t_{\rm{Nb}}=130\,\nano\meter$ thick Nb layer by magnetron sputtering and define the Nb microwave cavity by electron beam lithography (EBL) and reactive ion etching (RIE). Finally, we pattern the nanobeam with a length $L=60\,\micro\meter$ and a width $w=140\,\nano\meter$ by EBL and by using both anisotropic and isotropic etching to define and release the nanobeam (for details see Ref.~\cite{Zhou2013}). The effective mass of the center of mass coordinate of the in-plane vibration is $m_{\rm eff}\approx7\times10^{-15}\,\kilo\gram$. The nanomechanical oscillator is capacitively coupled to a $\lambda/4$ superconducting CPW microwave cavity with a resonance frequency of $\omega_{\rm c}/2\pi=6.07\,\giga\hertz$ and a quality factor of $Q\simeq8000$. An equivalent circuit diagram of the nano-electromechanical device is shown in the inset of Fig.~\ref{Hocke_Figure2}. The measurements are performed in a dilution fridge at a temperature of approximately $400\,\milli\kelvin$.

In our experiments we employ a setup similar to Refs.~\cite{Zhou2013, Gorodetsky2010a} to measure the amplitude fluctuations of the nanobeam (see Fig.\,\ref{Hocke_Figure2}). The mechanical motion is driven by the output signal of a vector network analyzer with angular frequency $\Omega_{\mathrm{ac}}$, which is applied directly to the sample via a bias tee. Mechanical spectroscopy is performed by analyzing the mechanically induced frequency fluctuations of the signal transmitted through the microwave cavity. To this end, we used a probe tone set to $\omega_{\rm p}=\omega_{\rm c}$ and a power of $P_{\mathrm{p}}=-83$\,dBm to minimize electromechanical back-action on the nanobeam. The probe tone $\omega_{\rm p}$  is frequency modulated with $\Omega_{\rm{mod}}/2\pi=\Omega_{\rm m}/2\pi-50\text{\,Hz}$ and a modulation depth of $100\,\kilo\hertz$. After transmission through the sample, the signal is amplified and downconverted with the frequency-modulated probe tone, corresponding to a homodyne detection. The phase quadrature is used as the input of the network analyzer. The frequency modulation experiences the same transduction as the mechanical motion (see Refs.~\cite{Zhou2013, Gorodetsky2010a} for further details). Together with a known temperature, this allows us to determine the a optomechanical coupling $g_0/(2\pi)=1.26\,\hertz$ and thus to calibrate the mechanical displacement spectrum.

\begin{figure}[tb]%
\center{\includegraphics[width=0.9\columnwidth]{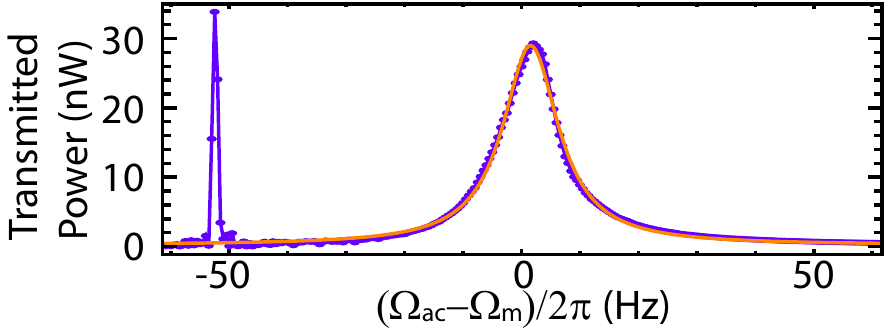}}
\caption{Transmitted microwave power spectrum measured by the vector network analyzer as a function of the drive frequency  $\Omega_{\mathrm{ac}}$ for an AC drive power of $P_{\mathrm{ac}}-91$\,dBm. The peak at $(\Omega_{\mathrm{ac}}-\Omega_{\mathrm{m}})/2\pi\simeq0\,\text{Hz}$ is caused by the motion of the nanobeam resulting in the  formation of mechanical sidebands on the transmitted microwave signal. The spectrum slightly deviates from a Lorentzian lineshape (orange line), indicating the onset of the Duffing response. Additionally, the calibration peak, originating form the frequency modulation of the probe tone is visible at $(\Omega_{\mathrm{ac}}-\Omega_{\mathrm{m}})/2\pi\simeq -50\,\text{Hz}$. }
\label{Hocke_Figure3}
\end{figure}

Figure~\ref{Hocke_Figure3} shows a typical spectrum as function of the drive frequency $\Omega_{\rm ac}$. The peak at a detuning of $-50\,\hertz$ stems from the frequency modulation of the probe tone $\omega_{\mathrm{p}}$ and is used for the calibration of the spectrum amplitude in terms of the microwave cavity frequency response. The central approximately Lorentzian peak is due to the mechanical motion of the nanobeam. Its slight deviation from the Lorentzian lineshape (solid orange line in Fig.\,\ref{Hocke_Figure3}) indicates the onset of nonlinearity.

\begin{figure}[tb]%
\center{\includegraphics[width=0.9\columnwidth]{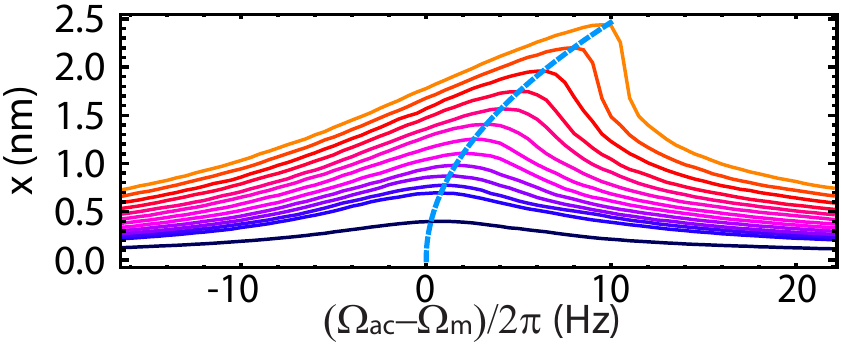}}
\caption{Measured mechanical displacement spectral density $S_x$ of the nanobeam plotted versus the drive frequency $\Omega_{\mathrm{ac}}$ for increasing driving powers from blue -100\,dBm (bottom curve). The other curves are obtained by increasing the drive power from $-95$\,dBm (blue) to $-84$\,dBm (orange) in 1\,dBm steps. The dotted light blue line is the backbone curve of the system obtained by fitting eq.~(\ref{equ:DuffingBackbone}) to the maximum amplitudes.  }
\label{Hocke_Figure4}
\end{figure}

In order to extract the stress $\sigma$ in the nanobeam, we first determine the mechanical eigenfrequency $\Omega_{\rm m}$ in the linear regime by recording the frequency spectrum at low drive power of $P_{\mathrm{ac}}=-100$\,dBm. At this power level we can neglect nonlinear effects (bottom black curve in Fig.~\ref{Hocke_Figure4}) and fit a Lorentzian to the data. Fitting the data yields  $\Omega_{\rm{m}}/2\pi\simeq1.45\,\mega\hertz$ and $\Gamma_{\rm m}/2\pi\simeq15\,\hertz$. With an effective mass density of $\rho_{\mathrm{eff}}=(\rho_{\mathrm{SiN}}t_{\mathrm{SiN}} + \rho_{\mathrm{Nb}}t_{\mathrm{Nb}})/(t_{\mathrm{SiN}} +t_{\mathrm{Nb}}) = 6621\,\kilo\gram/\meter^3$ we can use eq.~(\ref{equ:Sample_Om2}) to estimate the effective stress in the nanobeam to $\sigma_{\rm{eff}}=199\,\mega\pascal$. This agrees well with the value $\sigma_{\rm{eff}}=193\,\mega\pascal$ obtained from a bilayer beam theory (see supplemental material \cite{SI}), if we assume a prestressed tensile Si$_3$N$_4$ layer ($\sigma_{0,\mathrm{SiN}}=+830\,\mega\pascal$ \cite{Unterreithmeier2010a}) and a compressively pre-stressed Nb film ($\sigma_{0,\mathrm{Nb}}=-150\,\mega\pascal$) as determined from similar samples.

Next, we turn to the Duffing parameter. Increasing the AC drive power, we clearly observe a nonlinear behavior of the nanobeam. Fig.~\ref{Hocke_Figure4} shows the detected mechanical displacement spectrum for different AC drive powers. In all experiments the displacement spectral density was recorded for increasing drive frequency. The bottom curve is obtained for a drive power of  $P_{\mathrm{ac}}=-100$\,dBm, the remaining by increasing the drive power in $1$\,dB steps starting at $P_{\mathrm{ac}}=-95$\,dBm. The experimental data agree well with the expected evolution of a Duffing resonator. The dashed light blue line is a fit of the backbone curve according to eq.~(\ref{equ:DuffingBackbone}), yielding a Duffing parameter of $\alpha=1.99\times10^{11}\,\text{N/m}^3$. Note that Eq.~(\ref{equ:DuffingBackbone}) does not require any knowledge about the driving force applied to the nanobeam in the experiment. Thus it is sufficient, to determine the maximal mechanical amplitude $x_{\rm{0, max}}$ to extract $\alpha$ from the backbone curve. Plugging the value for $\alpha$  into (\ref{equ:DuffingParam}), we find a critical amplitude of $x_{\rm{crit}}= 2.57$\,nm.

With Eq.~(\ref{equ:Duffing_physical2}) we determine the effective Young's modulus $E_{\rm{eff}}$ using $\alpha=1.99\times10^{11}\,\text{N/m}^3$ and $\sigma_{\rm{eff}}=199$\,MPa and find $E_{\rm{eff}}=100$\,GPa for the investigated bilayer beam. We next compare this value to the prediction
\begin{equation}
E_{\mathrm{eff}} = \frac{E_{\mathrm{SiN}} t_{\mathrm{SiN}} + E_{\mathrm{Nb}} t_{\mathrm{Nb}}}{t_{\mathrm{SiN}}+t_{\mathrm{Nb}}}
\end{equation}
of the bilayer beam model (cf.~SI). Using the literature values $E_{\mathrm{SiN}}=160\,\giga\pascal$ \cite{Unterreithmeier2010a} and $E_{\mathrm{Nb}}=105\,\giga\pascal$ \cite{cleland_foundations_2003}, we get $E_{\mathrm{eff}}=124\,\giga\pascal$. This value agrees well with the experimental value of $E_{\rm{eff}}=100$\,GPa.

In summary, we have presented a method to quantitatively determine the effective mechanical parameters of a bilayer nanobeam embedded into an electromechanical system. Our method is based on the analysis of the oscillation amplitude of a driven nanomechanical beam both in the linear and non-linear regime described by the Duffing equation. Measuring the amplitude spectrum of the beam as a function of the drive power allows us to extract the effective prestress and effective Young's modulus of the beam. Our results show that the spectroscopic analysis of the mechanical eigenfrequency in the linear and nonlinear regime is a powerful tool to characterize mechanical oscillators fabricated from more than one layer. We also have developed an analytical model description of the mechanical behavior of pre-stressed nanomechanical beams consisting of two (or more) layers. The model predictions agree well with the experiment. This demonstrates that this model can be used to tailor and optimize the mechanical parameters of multi-layer nanomechanical beams.

T.J.K. acknowledges support by the ERC grant SIMP and X.Z. by the NCCR of Quantum Engineering.

%\bibliography{bibliography-Duffing}

\begin{thebibliography}{36}%
\makeatletter
\providecommand \@ifxundefined [1]{%
 \@ifx{#1\undefined}
}%
\providecommand \@ifnum [1]{%
 \ifnum #1\expandafter \@firstoftwo
 \else \expandafter \@secondoftwo
 \fi
}%
\providecommand \@ifx [1]{%
 \ifx #1\expandafter \@firstoftwo
 \else \expandafter \@secondoftwo
 \fi
}%
\providecommand \natexlab [1]{#1}%
\providecommand \enquote  [1]{``#1''}%
\providecommand \bibnamefont  [1]{#1}%
\providecommand \bibfnamefont [1]{#1}%
\providecommand \citenamefont [1]{#1}%
\providecommand \href@noop [0]{\@secondoftwo}%
\providecommand \href [0]{\begingroup \@sanitize@url \@href}%
\providecommand \@href[1]{\@@startlink{#1}\@@href}%
\providecommand \@@href[1]{\endgroup#1\@@endlink}%
\providecommand \@sanitize@url [0]{\catcode `\\12\catcode `\$12\catcode
  `\&12\catcode `\#12\catcode `\^12\catcode `\_12\catcode `\%12\relax}%
\providecommand \@@startlink[1]{}%
\providecommand \@@endlink[0]{}%
\providecommand \url  [0]{\begingroup\@sanitize@url \@url }%
\providecommand \@url [1]{\endgroup\@href {#1}{\urlprefix }}%
\providecommand \urlprefix  [0]{URL }%
\providecommand \Eprint [0]{\href }%
\providecommand \doibase [0]{http://dx.doi.org/}%
\providecommand \selectlanguage [0]{\@gobble}%
\providecommand \bibinfo  [0]{\@secondoftwo}%
\providecommand \bibfield  [0]{\@secondoftwo}%
\providecommand \translation [1]{[#1]}%
\providecommand \BibitemOpen [0]{}%
\providecommand \bibitemStop [0]{}%
\providecommand \bibitemNoStop [0]{.\EOS\space}%
\providecommand \EOS [0]{\spacefactor3000\relax}%
\providecommand \BibitemShut  [1]{\csname bibitem#1\endcsname}%
\let\auto@bib@innerbib\@empty
%</preamble>
\bibitem [{\citenamefont {Masmanidis}\ \emph {et~al.}(2005)\citenamefont
  {Masmanidis}, \citenamefont {Tang}, \citenamefont {Myers}, \citenamefont
  {Li}, \citenamefont {De~Greve}, \citenamefont {Vermeulen}, \citenamefont
  {Van Roy},\ and\ \citenamefont {Roukes}}]{Masmanidis2005}%
  \BibitemOpen
  \bibfield  {author} {\bibinfo {author} {\bibfnamefont {S.~C.}\ \bibnamefont
  {Masmanidis}}, \bibinfo {author} {\bibfnamefont {H.~X.}\ \bibnamefont
  {Tang}}, \bibinfo {author} {\bibfnamefont {E.~B.}\ \bibnamefont {Myers}},
  \bibinfo {author} {\bibfnamefont {M.}~\bibnamefont {Li}}, \bibinfo {author}
  {\bibfnamefont {K.}~\bibnamefont {De~Greve}}, \bibinfo {author}
  {\bibfnamefont {G.}~\bibnamefont {Vermeulen}}, \bibinfo {author}
  {\bibfnamefont {W.}~\bibnamefont {Van Roy}}, \ and\ \bibinfo {author}
  {\bibfnamefont {M.~L.}\ \bibnamefont {Roukes}},\ }\href {\doibase
  10.1103/PhysRevLett.95.187206} {\bibfield  {journal} {\bibinfo  {journal}
  {Phys. Rev. Lett.}\ }\textbf {\bibinfo {volume} {95}},\ \bibinfo {pages}
  {187206} (\bibinfo {year} {2005})}\BibitemShut {NoStop}%
\bibitem [{\citenamefont {Feng}\ \emph {et~al.}(2008)\citenamefont {Feng},
  \citenamefont {White}, \citenamefont {Hajimiri},\ and\ \citenamefont
  {Roukes}}]{Feng:2008a}%
  \BibitemOpen
  \bibfield  {author} {\bibinfo {author} {\bibfnamefont {X.~L.}\ \bibnamefont
  {Feng}}, \bibinfo {author} {\bibfnamefont {C.~J.}\ \bibnamefont {White}},
  \bibinfo {author} {\bibfnamefont {A.}~\bibnamefont {Hajimiri}}, \ and\
  \bibinfo {author} {\bibfnamefont {M.~L.}\ \bibnamefont {Roukes}},\ }\href
  {http://dx.doi.org/10.1038/nnano.2008.125} {\bibfield  {journal} {\bibinfo
  {journal} {Nat. Nano}\ }\textbf {\bibinfo {volume} {3}},\ \bibinfo {pages}
  {342--346} (\bibinfo {year} {2008})}\BibitemShut {NoStop}%
\bibitem [{\citenamefont {Zolfagharkhani}\ \emph {et~al.}(2008)\citenamefont
  {Zolfagharkhani}, \citenamefont {Gaidarzhy}, \citenamefont {Degiovanni},
  \citenamefont {Kettemann}, \citenamefont {Fulde},\ and\ \citenamefont
  {Mohanty}}]{Zolfagharkhani2008}%
  \BibitemOpen
  \bibfield  {author} {\bibinfo {author} {\bibfnamefont {G.}~\bibnamefont
  {Zolfagharkhani}}, \bibinfo {author} {\bibfnamefont {A.}~\bibnamefont
  {Gaidarzhy}}, \bibinfo {author} {\bibfnamefont {P.}~\bibnamefont
  {Degiovanni}}, \bibinfo {author} {\bibfnamefont {S.}~\bibnamefont
  {Kettemann}}, \bibinfo {author} {\bibfnamefont {P.}~\bibnamefont {Fulde}}, \
  and\ \bibinfo {author} {\bibfnamefont {P.}~\bibnamefont {Mohanty}},\ }\href
  {\doibase 10.1038/nnano.2008.311} {\bibfield  {journal} {\bibinfo  {journal}
  {Nat. Nano}\ }\textbf {\bibinfo {volume} {3}},\ \bibinfo {pages} {720--723}
  (\bibinfo {year} {2008})}\BibitemShut {NoStop}%
\bibitem [{\citenamefont {Naik}\ \emph {et~al.}(2009)\citenamefont {Naik},
  \citenamefont {Hanay}, \citenamefont {Hiebert}, \citenamefont {Feng},\ and\
  \citenamefont {Roukes}}]{Naik:2009a}%
  \BibitemOpen
  \bibfield  {author} {\bibinfo {author} {\bibfnamefont {A.~K.}\ \bibnamefont
  {Naik}}, \bibinfo {author} {\bibfnamefont {M.~S.}\ \bibnamefont {Hanay}},
  \bibinfo {author} {\bibfnamefont {W.~K.}\ \bibnamefont {Hiebert}}, \bibinfo
  {author} {\bibfnamefont {X.~L.}\ \bibnamefont {Feng}}, \ and\ \bibinfo
  {author} {\bibfnamefont {M.~L.}\ \bibnamefont {Roukes}},\ }\bibfield  {title}
  {\enquote {\bibinfo {title} {Towards single-molecule nanomechanical mass
  spectrometry},}\ }\href {\doibase 10.1038/nnano.2009.152} {\bibfield
  {journal} {\bibinfo  {journal} {Nat. Nano}\ }\textbf {\bibinfo {volume}
  {4}},\ \bibinfo {pages} {445--450} (\bibinfo {year} {2009})}\BibitemShut
  {NoStop}%
\bibitem [{\citenamefont {Chaste}\ \emph {et~al.}(2012)\citenamefont {Chaste},
  \citenamefont {Eichler}, \citenamefont {Moser}, \citenamefont {Ceballos},
  \citenamefont {Rurali},\ and\ \citenamefont {Bachtold}}]{Chaste2011}%
  \BibitemOpen
  \bibfield  {author} {\bibinfo {author} {\bibfnamefont {J.}~\bibnamefont
  {Chaste}}, \bibinfo {author} {\bibfnamefont {A.}~\bibnamefont {Eichler}},
  \bibinfo {author} {\bibfnamefont {J.}~\bibnamefont {Moser}}, \bibinfo
  {author} {\bibfnamefont {G.}~\bibnamefont {Ceballos}}, \bibinfo {author}
  {\bibfnamefont {R.}~\bibnamefont {Rurali}}, \ and\ \bibinfo {author}
  {\bibfnamefont {A.}~\bibnamefont {Bachtold}},\ }\href {\doibase
  10.1038/nnano.2012.42} {\bibfield  {journal} {\bibinfo  {journal} {Nat.
  Nanotech.}\ }\textbf {\bibinfo {volume} {7}},\ \bibinfo {pages} {301--304}
  (\bibinfo {year} {2012})}\BibitemShut {NoStop}%
\bibitem [{\citenamefont {Hanay}\ \emph {et~al.}(2012)\citenamefont {Hanay},
  \citenamefont {Kelber}, \citenamefont {Naik}, \citenamefont {Chi},
  \citenamefont {Hentz}, \citenamefont {Bullard}, \citenamefont {Colinet},
  \citenamefont {Duraffourg},\ and\ \citenamefont {Roukes}}]{Hanay:2012a}%
  \BibitemOpen
  \bibfield  {author} {\bibinfo {author} {\bibfnamefont {M.~S.}\ \bibnamefont
  {Hanay}}, \bibinfo {author} {\bibfnamefont {S.}~\bibnamefont {Kelber}},
  \bibinfo {author} {\bibfnamefont {A.~K.}\ \bibnamefont {Naik}}, \bibinfo
  {author} {\bibfnamefont {D.}~\bibnamefont {Chi}}, \bibinfo {author}
  {\bibfnamefont {S.}~\bibnamefont {Hentz}}, \bibinfo {author} {\bibfnamefont
  {C.~E.}\ \bibnamefont {Bullard}}, \bibinfo {author} {\bibfnamefont
  {E.}~\bibnamefont {Colinet}}, \bibinfo {author} {\bibfnamefont
  {L.}~\bibnamefont {Duraffourg}}, \ and\ \bibinfo {author} {\bibfnamefont
  {M.~L.}\ \bibnamefont {Roukes}},\ }\href {\doibase 10.1038/nnano.2012.119}
  {\bibfield  {journal} {\bibinfo  {journal} {Nat. Nano}\ }\textbf {\bibinfo
  {volume} {7}},\ \bibinfo {pages} {602--608} (\bibinfo {year}
  {2012})}\BibitemShut {NoStop}%
\bibitem [{\citenamefont {Aspelmeyer}, \citenamefont {Kippenberg},\ and\
  \citenamefont {Marquardt}(2013)}]{Aspelmeyer2013}%
  \BibitemOpen
  \bibfield  {author} {\bibinfo {author} {\bibfnamefont {M.}~\bibnamefont
  {Aspelmeyer}}, \bibinfo {author} {\bibfnamefont {T.~J.}\ \bibnamefont
  {Kippenberg}}, \ and\ \bibinfo {author} {\bibfnamefont {F.}~\bibnamefont
  {Marquardt}},\ }\href {http://arxiv.org/abs/1303.0733} {\bibfield  {journal}
  {\bibinfo  {journal} {ArXiv:1303.0733}\ } (\bibinfo {year}
  {2013})}\BibitemShut {NoStop}%
\bibitem [{\citenamefont {Aspelmeyer}, \citenamefont {Kippenberg},\ and\
  \citenamefont {Marquardt}(2014)}]{aspelmeyer_cavity_2014}%
  \BibitemOpen
  \bibinfo {editor} {\bibfnamefont {M.}~\bibnamefont {Aspelmeyer}}, \bibinfo
  {editor} {\bibfnamefont {T.~J.}\ \bibnamefont {Kippenberg}}, \ and\ \bibinfo
  {editor} {\bibfnamefont {F.}~\bibnamefont {Marquardt}},\ eds.,\ \href@noop {}
  {\emph {\bibinfo {title} {Cavity Optomechanics - Nano- and Micromechanical
  Resonators Interacting with Light}}}\ (\bibinfo  {publisher} {Springer},\
  \bibinfo {year} {2014})\BibitemShut {NoStop}%
\bibitem [{\citenamefont {Regal}, \citenamefont {Teufel},\ and\ \citenamefont
  {Lehnert}(2008)}]{Regal2008}%
  \BibitemOpen
  \bibfield  {author} {\bibinfo {author} {\bibfnamefont {C.~A.}\ \bibnamefont
  {Regal}}, \bibinfo {author} {\bibfnamefont {J.~D.}\ \bibnamefont {Teufel}}, \
  and\ \bibinfo {author} {\bibfnamefont {K.~W.}\ \bibnamefont {Lehnert}},\
  }\href {http://dx.doi.org/10.1038/nphys974} {\bibfield  {journal} {\bibinfo
  {journal} {Nat. Phys.}\ }\textbf {\bibinfo {volume} {4}},\ \bibinfo {pages}
  {555--560} (\bibinfo {year} {2008})}\BibitemShut {NoStop}%
\bibitem [{\citenamefont {Hertzberg}\ \emph {et~al.}(2009)\citenamefont
  {Hertzberg}, \citenamefont {Rocheleau}, \citenamefont {Ndukum}, \citenamefont
  {Savva}, \citenamefont {Clerk},\ and\ \citenamefont
  {Schwab}}]{Hertzberg2009a}%
  \BibitemOpen
  \bibfield  {author} {\bibinfo {author} {\bibfnamefont {J.~B.}\ \bibnamefont
  {Hertzberg}}, \bibinfo {author} {\bibfnamefont {T.}~\bibnamefont
  {Rocheleau}}, \bibinfo {author} {\bibfnamefont {T.}~\bibnamefont {Ndukum}},
  \bibinfo {author} {\bibfnamefont {M.}~\bibnamefont {Savva}}, \bibinfo
  {author} {\bibfnamefont {A.~A.}\ \bibnamefont {Clerk}}, \ and\ \bibinfo
  {author} {\bibfnamefont {K.~C.}\ \bibnamefont {Schwab}},\ }\href
  {http://dx.doi.org/10.1038/nphys1479} {\bibfield  {journal} {\bibinfo
  {journal} {Nat. Phys.}\ }\textbf {\bibinfo {volume} {6}},\ \bibinfo {pages}
  {213--217} (\bibinfo {year} {2009})}\BibitemShut {NoStop}%
\bibitem [{\citenamefont {Massel}\ \emph {et~al.}(2011)\citenamefont {Massel},
  \citenamefont {Heikkila\"{a}}, \citenamefont {Pirkkalainen}, \citenamefont
  {Cho}, \citenamefont {Saloniemi}, \citenamefont {Hakonen},\ and\
  \citenamefont {Sillanp\"{a}\"{a}}}]{Massel2011}%
  \BibitemOpen
  \bibfield  {author} {\bibinfo {author} {\bibfnamefont {F.}~\bibnamefont
  {Massel}}, \bibinfo {author} {\bibfnamefont {T.~T.}\ \bibnamefont
  {Heikkila\"{a}}}, \bibinfo {author} {\bibfnamefont {J.-M.}\ \bibnamefont
  {Pirkkalainen}}, \bibinfo {author} {\bibfnamefont {S.~U.}\ \bibnamefont
  {Cho}}, \bibinfo {author} {\bibfnamefont {H.}~\bibnamefont {Saloniemi}},
  \bibinfo {author} {\bibfnamefont {P.~J.}\ \bibnamefont {Hakonen}}, \ and\
  \bibinfo {author} {\bibfnamefont {M.~A.}\ \bibnamefont {Sillanp\"{a}\"{a}}},\
  }\href {\doibase 10.1038/nature10628} {\bibfield  {journal} {\bibinfo
  {journal} {Nature}\ }\textbf {\bibinfo {volume} {480}},\ \bibinfo {pages}
  {351--354} (\bibinfo {year} {2011})}\BibitemShut {NoStop}%
\bibitem [{\citenamefont {Teufel}\ \emph
  {et~al.}(2011{\natexlab{a}})\citenamefont {Teufel}, \citenamefont {Li},
  \citenamefont {Allman}, \citenamefont {Cicak}, \citenamefont {Sirois},
  \citenamefont {Whittaker},\ and\ \citenamefont {Simmonds}}]{Teufel2011}%
  \BibitemOpen
  \bibfield  {author} {\bibinfo {author} {\bibfnamefont {J.~D.}\ \bibnamefont
  {Teufel}}, \bibinfo {author} {\bibfnamefont {D.}~\bibnamefont {Li}}, \bibinfo
  {author} {\bibfnamefont {M.~S.}\ \bibnamefont {Allman}}, \bibinfo {author}
  {\bibfnamefont {K.}~\bibnamefont {Cicak}}, \bibinfo {author} {\bibfnamefont
  {A.~J.}\ \bibnamefont {Sirois}}, \bibinfo {author} {\bibfnamefont {J.~D.}\
  \bibnamefont {Whittaker}}, \ and\ \bibinfo {author} {\bibfnamefont {R.~W.}\
  \bibnamefont {Simmonds}},\ }\href {\doibase 10.1038/nature09898} {\bibfield
  {journal} {\bibinfo  {journal} {Nature}\ }\textbf {\bibinfo {volume} {471}},\
  \bibinfo {pages} {204--208} (\bibinfo {year}
  {2011}{\natexlab{a}})}\BibitemShut {NoStop}%
\bibitem [{\citenamefont {Teufel}\ \emph
  {et~al.}(2011{\natexlab{b}})\citenamefont {Teufel}, \citenamefont {Donner},
  \citenamefont {Li}, \citenamefont {Harlow}, \citenamefont {Allman},
  \citenamefont {Cicak}, \citenamefont {Sirois}, \citenamefont {Whittaker},
  \citenamefont {Lehnert},\ and\ \citenamefont {Simmonds}}]{Teufel2011a}%
  \BibitemOpen
  \bibfield  {author} {\bibinfo {author} {\bibfnamefont {J.~D.}\ \bibnamefont
  {Teufel}}, \bibinfo {author} {\bibfnamefont {T.}~\bibnamefont {Donner}},
  \bibinfo {author} {\bibfnamefont {D.}~\bibnamefont {Li}}, \bibinfo {author}
  {\bibfnamefont {J.~W.}\ \bibnamefont {Harlow}}, \bibinfo {author}
  {\bibfnamefont {M.~S.}\ \bibnamefont {Allman}}, \bibinfo {author}
  {\bibfnamefont {K.}~\bibnamefont {Cicak}}, \bibinfo {author} {\bibfnamefont
  {A.~J.}\ \bibnamefont {Sirois}}, \bibinfo {author} {\bibfnamefont {J.~D.}\
  \bibnamefont {Whittaker}}, \bibinfo {author} {\bibfnamefont {K.~W.}\
  \bibnamefont {Lehnert}}, \ and\ \bibinfo {author} {\bibfnamefont {R.~W.}\
  \bibnamefont {Simmonds}},\ }\href {\doibase 10.1038/nature10261} {\bibfield
  {journal} {\bibinfo  {journal} {Nature}\ }\textbf {\bibinfo {volume} {475}},\
  \bibinfo {pages} {359--363} (\bibinfo {year}
  {2011}{\natexlab{b}})}\BibitemShut {NoStop}%
\bibitem [{\citenamefont {Hocke}\ \emph {et~al.}(2012)\citenamefont {Hocke},
  \citenamefont {Zhou}, \citenamefont {Schliesser}, \citenamefont {Kippenberg},
  \citenamefont {Huebl},\ and\ \citenamefont {Gross}}]{Hocke2012}%
  \BibitemOpen
  \bibfield  {author} {\bibinfo {author} {\bibfnamefont {F.}~\bibnamefont
  {Hocke}}, \bibinfo {author} {\bibfnamefont {X.}~\bibnamefont {Zhou}},
  \bibinfo {author} {\bibfnamefont {A.}~\bibnamefont {Schliesser}}, \bibinfo
  {author} {\bibfnamefont {T.~J.}\ \bibnamefont {Kippenberg}}, \bibinfo
  {author} {\bibfnamefont {H.}~\bibnamefont {Huebl}}, \ and\ \bibinfo {author}
  {\bibfnamefont {R.}~\bibnamefont {Gross}},\ }\href {\doibase
  10.1088/1367-2630/14/12/123037} {\bibfield  {journal} {\bibinfo  {journal}
  {N. J. Phys.}\ }\textbf {\bibinfo {volume} {14}},\ \bibinfo {pages}
  {123037--} (\bibinfo {year} {2012})}\BibitemShut {NoStop}%
\bibitem [{\citenamefont {Zhou}\ \emph {et~al.}(2013)\citenamefont {Zhou},
  \citenamefont {Hocke}, \citenamefont {Schliesser}, \citenamefont {Marx},
  \citenamefont {Huebl}, \citenamefont {Gross},\ and\ \citenamefont
  {Kippenberg}}]{Zhou2013}%
  \BibitemOpen
  \bibfield  {author} {\bibinfo {author} {\bibfnamefont {X.}~\bibnamefont
  {Zhou}}, \bibinfo {author} {\bibfnamefont {F.}~\bibnamefont {Hocke}},
  \bibinfo {author} {\bibfnamefont {A.}~\bibnamefont {Schliesser}}, \bibinfo
  {author} {\bibfnamefont {A.}~\bibnamefont {Marx}}, \bibinfo {author}
  {\bibfnamefont {H.}~\bibnamefont {Huebl}}, \bibinfo {author} {\bibfnamefont
  {R.}~\bibnamefont {Gross}}, \ and\ \bibinfo {author} {\bibfnamefont {T.~J.}\
  \bibnamefont {Kippenberg}},\ }\href {\doibase 10.1038/nphys2527} {\bibfield
  {journal} {\bibinfo  {journal} {Nat. Phys.}\ }\textbf {\bibinfo {volume}
  {9}},\ \bibinfo {pages} {179--184} (\bibinfo {year} {2013})}\BibitemShut
  {NoStop}%
\bibitem [{\citenamefont {Palomaki}\ \emph {et~al.}(2013)\citenamefont
  {Palomaki}, \citenamefont {Harlow}, \citenamefont {Teufel}, \citenamefont
  {Simmonds},\ and\ \citenamefont {Lehnert}}]{Palomaki2012}%
  \BibitemOpen
  \bibfield  {author} {\bibinfo {author} {\bibfnamefont {T.~A.}\ \bibnamefont
  {Palomaki}}, \bibinfo {author} {\bibfnamefont {J.~W.}\ \bibnamefont
  {Harlow}}, \bibinfo {author} {\bibfnamefont {J.~D.}\ \bibnamefont {Teufel}},
  \bibinfo {author} {\bibfnamefont {R.~W.}\ \bibnamefont {Simmonds}}, \ and\
  \bibinfo {author} {\bibfnamefont {K.~W.}\ \bibnamefont {Lehnert}},\ }\href
  {\doibase 10.1038/nature11915} {\bibfield  {journal} {\bibinfo  {journal}
  {Nature}\ }\textbf {\bibinfo {volume} {495}},\ \bibinfo {pages} {210--214}
  (\bibinfo {year} {2013})}\BibitemShut {NoStop}%
\bibitem [{\citenamefont {Sulkko}\ \emph {et~al.}(2010)\citenamefont {Sulkko},
  \citenamefont {Sillanp\"{a}\"{a}}, \citenamefont {H\"{a}kkinen},
  \citenamefont {Lechner}, \citenamefont {Helle}, \citenamefont {Fefferman},
  \citenamefont {Parpia},\ and\ \citenamefont {Hakonen}}]{Sulkko2010}%
  \BibitemOpen
  \bibfield  {author} {\bibinfo {author} {\bibfnamefont {J.}~\bibnamefont
  {Sulkko}}, \bibinfo {author} {\bibfnamefont {M.~A.}\ \bibnamefont
  {Sillanp\"{a}\"{a}}}, \bibinfo {author} {\bibfnamefont {P.}~\bibnamefont
  {H\"{a}kkinen}}, \bibinfo {author} {\bibfnamefont {L.}~\bibnamefont
  {Lechner}}, \bibinfo {author} {\bibfnamefont {M.}~\bibnamefont {Helle}},
  \bibinfo {author} {\bibfnamefont {A.}~\bibnamefont {Fefferman}}, \bibinfo
  {author} {\bibfnamefont {J.}~\bibnamefont {Parpia}}, \ and\ \bibinfo {author}
  {\bibfnamefont {P.~J.}\ \bibnamefont {Hakonen}},\ }\href
  {http://dx.doi.org/10.1021/nl102771p} {\bibfield  {journal} {\bibinfo
  {journal} {Nano Lett.}\ }\textbf {\bibinfo {volume} {10}},\ \bibinfo {pages}
  {4884--4889} (\bibinfo {year} {2010})}\BibitemShut {NoStop}%
\bibitem [{\citenamefont {Hoch}\ \emph {et~al.}(2011)\citenamefont {Hoch},
  \citenamefont {Montague}, \citenamefont {Bright}, \citenamefont {Rogers},
  \citenamefont {Bertness}, \citenamefont {Teufel},\ and\ \citenamefont
  {Lehnert}}]{Hoch2011}%
  \BibitemOpen
  \bibfield  {author} {\bibinfo {author} {\bibfnamefont {S.~W.}\ \bibnamefont
  {Hoch}}, \bibinfo {author} {\bibfnamefont {J.~R.}\ \bibnamefont {Montague}},
  \bibinfo {author} {\bibfnamefont {V.~M.}\ \bibnamefont {Bright}}, \bibinfo
  {author} {\bibfnamefont {C.~T.}\ \bibnamefont {Rogers}}, \bibinfo {author}
  {\bibfnamefont {K.~A.}\ \bibnamefont {Bertness}}, \bibinfo {author}
  {\bibfnamefont {J.~D.}\ \bibnamefont {Teufel}}, \ and\ \bibinfo {author}
  {\bibfnamefont {K.~W.}\ \bibnamefont {Lehnert}},\ }\href {\doibase
  http://dx.doi.org/10.1063/1.3614562} {\bibfield  {journal} {\bibinfo
  {journal} {Appl. Phys. Lett.}\ }\textbf {\bibinfo {volume} {99}},\ \bibinfo
  {pages} {053101} (\bibinfo {year} {2011})},\ \bibinfo {note} {journal
  article}\BibitemShut {NoStop}%
\bibitem [{\citenamefont {Faust}\ \emph {et~al.}(2012)\citenamefont {Faust},
  \citenamefont {Rieger}, \citenamefont {Seitner}, \citenamefont {Krenn},
  \citenamefont {Kotthaus},\ and\ \citenamefont {Weig}}]{Faust2012a}%
  \BibitemOpen
  \bibfield  {author} {\bibinfo {author} {\bibfnamefont {T.}~\bibnamefont
  {Faust}}, \bibinfo {author} {\bibfnamefont {J.}~\bibnamefont {Rieger}},
  \bibinfo {author} {\bibfnamefont {M.~J.}\ \bibnamefont {Seitner}}, \bibinfo
  {author} {\bibfnamefont {P.}~\bibnamefont {Krenn}}, \bibinfo {author}
  {\bibfnamefont {J.~P.}\ \bibnamefont {Kotthaus}}, \ and\ \bibinfo {author}
  {\bibfnamefont {E.~M.}\ \bibnamefont {Weig}},\ }\href {\doibase
  10.1103/PhysRevLett.109.037205} {\bibfield  {journal} {\bibinfo  {journal}
  {Phys. Rev. Lett.}\ }\textbf {\bibinfo {volume} {109}},\ \bibinfo {pages}
  {037205} (\bibinfo {year} {2012})}\BibitemShut {NoStop}%
\bibitem [{\citenamefont {Sattler}(2011)}]{Sattler2011}%
  \BibitemOpen
  \bibinfo {editor} {\bibfnamefont {K.~D.}\ \bibnamefont {Sattler}},\ ed.,\
  \enquote {\bibinfo {title} {Handbook of nanophysics: Functional
  nanomaterials},}\ \ (\bibinfo  {publisher} {CRC press: Boca Raton},\ \bibinfo
  {year} {2011})\ Chap.~\bibinfo {chapter} {8}\BibitemShut {NoStop}%
\bibitem [{\citenamefont {Kozinsky}\ \emph {et~al.}(2006)\citenamefont
  {Kozinsky}, \citenamefont {Postma}, \citenamefont {Bargatin},\ and\
  \citenamefont {Roukes}}]{Kozinsky2006}%
  \BibitemOpen
  \bibfield  {author} {\bibinfo {author} {\bibfnamefont {I.}~\bibnamefont
  {Kozinsky}}, \bibinfo {author} {\bibfnamefont {H.~W.~C.}\ \bibnamefont
  {Postma}}, \bibinfo {author} {\bibfnamefont {I.}~\bibnamefont {Bargatin}}, \
  and\ \bibinfo {author} {\bibfnamefont {M.~L.}\ \bibnamefont {Roukes}},\
  }\href {\doibase http://dx.doi.org/10.1063/1.2209211} {\bibfield  {journal}
  {\bibinfo  {journal} {Appl. Phys. Lett.}\ }\textbf {\bibinfo {volume} {88}},\
  \bibinfo {pages} {253101--3} (\bibinfo {year} {2006})}\BibitemShut {NoStop}%
\bibitem [{\citenamefont {Kozinsky}\ \emph {et~al.}(2007)\citenamefont
  {Kozinsky}, \citenamefont {Postma}, \citenamefont {Kogan}, \citenamefont
  {Husain},\ and\ \citenamefont {Roukes}}]{Kozinsky2007}%
  \BibitemOpen
  \bibfield  {author} {\bibinfo {author} {\bibfnamefont {I.}~\bibnamefont
  {Kozinsky}}, \bibinfo {author} {\bibfnamefont {H.~W.~C.}\ \bibnamefont
  {Postma}}, \bibinfo {author} {\bibfnamefont {O.}~\bibnamefont {Kogan}},
  \bibinfo {author} {\bibfnamefont {A.}~\bibnamefont {Husain}}, \ and\ \bibinfo
  {author} {\bibfnamefont {M.~L.}\ \bibnamefont {Roukes}},\ }\href {\doibase
  10.1103/PhysRevLett.99.207201} {\bibfield  {journal} {\bibinfo  {journal}
  {Phys. Rev. Lett.}\ }\textbf {\bibinfo {volume} {99}},\ \bibinfo {pages}
  {207201} (\bibinfo {year} {2007})}\BibitemShut {NoStop}%
\bibitem [{\citenamefont {Karabalin}\ \emph {et~al.}(2009)\citenamefont
  {Karabalin}, \citenamefont {Matheny}, \citenamefont {Feng}, \citenamefont
  {Defay}, \citenamefont {Le~Rhun}, \citenamefont {Marcoux}, \citenamefont
  {Hentz}, \citenamefont {Andreucci},\ and\ \citenamefont
  {Roukes}}]{Karabalin2009a}%
  \BibitemOpen
  \bibfield  {author} {\bibinfo {author} {\bibfnamefont {R.~B.}\ \bibnamefont
  {Karabalin}}, \bibinfo {author} {\bibfnamefont {M.~H.}\ \bibnamefont
  {Matheny}}, \bibinfo {author} {\bibfnamefont {X.~L.}\ \bibnamefont {Feng}},
  \bibinfo {author} {\bibfnamefont {E.}~\bibnamefont {Defay}}, \bibinfo
  {author} {\bibfnamefont {G.}~\bibnamefont {Le~Rhun}}, \bibinfo {author}
  {\bibfnamefont {C.}~\bibnamefont {Marcoux}}, \bibinfo {author} {\bibfnamefont
  {S.}~\bibnamefont {Hentz}}, \bibinfo {author} {\bibfnamefont
  {P.}~\bibnamefont {Andreucci}}, \ and\ \bibinfo {author} {\bibfnamefont
  {M.~L.}\ \bibnamefont {Roukes}},\ }\href {\doibase 10.1063/1.3216586}
  {\bibfield  {journal} {\bibinfo  {journal} {Appl. Phys. Lett.}\ }\textbf
  {\bibinfo {volume} {95}},\ \bibinfo {pages} {103111--3} (\bibinfo {year}
  {2009})}\BibitemShut {NoStop}%
\bibitem [{\citenamefont {Unterreithmeier}, \citenamefont {Manus},\ and\
  \citenamefont {Kotthaus}(2009)}]{Unterreithmeier2009a}%
  \BibitemOpen
  \bibfield  {author} {\bibinfo {author} {\bibfnamefont {Q.~P.}\ \bibnamefont
  {Unterreithmeier}}, \bibinfo {author} {\bibfnamefont {S.}~\bibnamefont
  {Manus}}, \ and\ \bibinfo {author} {\bibfnamefont {J.~P.}\ \bibnamefont
  {Kotthaus}},\ }\bibfield  {title} {\enquote {\bibinfo {title} {Coherent
  detection of nonlinear nanomechanical motion using a stroboscopic
  downconversion technique},}\ }\href {\doibase 10.1063/1.3155164} {\bibfield
  {journal} {\bibinfo  {journal} {Appl. Phys. Lett.}\ }\textbf {\bibinfo
  {volume} {94}},\ \bibinfo {pages} {263104--3} (\bibinfo {year}
  {2009})}\BibitemShut {NoStop}%
\bibitem [{\citenamefont {Venkatesan}\ \emph {et~al.}(2010)\citenamefont
  {Venkatesan}, \citenamefont {Lulla}, \citenamefont {Patton}, \citenamefont
  {Armour}, \citenamefont {Mellor},\ and\ \citenamefont
  {Owers-Bradley}}]{Venkatesan2010}%
  \BibitemOpen
  \bibfield  {author} {\bibinfo {author} {\bibfnamefont {A.}~\bibnamefont
  {Venkatesan}}, \bibinfo {author} {\bibfnamefont {K.~J.}\ \bibnamefont
  {Lulla}}, \bibinfo {author} {\bibfnamefont {M.~J.}\ \bibnamefont {Patton}},
  \bibinfo {author} {\bibfnamefont {A.~D.}\ \bibnamefont {Armour}}, \bibinfo
  {author} {\bibfnamefont {C.~J.}\ \bibnamefont {Mellor}}, \ and\ \bibinfo
  {author} {\bibfnamefont {J.~R.}\ \bibnamefont {Owers-Bradley}},\ }\href
  {\doibase 10.1103/PhysRevB.81.073410} {\bibfield  {journal} {\bibinfo
  {journal} {Phys. Rev. B}\ }\textbf {\bibinfo {volume} {81}},\ \bibinfo
  {pages} {073410--} (\bibinfo {year} {2010})}\BibitemShut {NoStop}%
\bibitem [{\citenamefont {Juillard}\ \emph {et~al.}(2010)\citenamefont
  {Juillard}, \citenamefont {Bonnoit}, \citenamefont {Avignon}, \citenamefont
  {Hentz},\ and\ \citenamefont {Colinet}}]{Juillard2010}%
  \BibitemOpen
  \bibfield  {author} {\bibinfo {author} {\bibfnamefont {J.}~\bibnamefont
  {Juillard}}, \bibinfo {author} {\bibfnamefont {A.}~\bibnamefont {Bonnoit}},
  \bibinfo {author} {\bibfnamefont {E.}~\bibnamefont {Avignon}}, \bibinfo
  {author} {\bibfnamefont {S.}~\bibnamefont {Hentz}}, \ and\ \bibinfo {author}
  {\bibfnamefont {E.}~\bibnamefont {Colinet}},\ }\href {\doibase
  10.1063/1.3277022} {\bibfield  {journal} {\bibinfo  {journal} {J. Appl.
  Phys.}\ }\textbf {\bibinfo {volume} {107}},\ \bibinfo {pages} {014907--10}
  (\bibinfo {year} {2010})}\BibitemShut {NoStop}%
\bibitem [{\citenamefont {Jun}\ \emph {et~al.}(2010)\citenamefont {Jun},
  \citenamefont {Moon}, \citenamefont {Kim}, \citenamefont {Cho}, \citenamefont
  {Kang}, \citenamefont {Jung}, \citenamefont {Yoon}, \citenamefont {Shin},
  \citenamefont {Song}, \citenamefont {Choi}, \citenamefont {Choi},
  \citenamefont {Bae}, \citenamefont {Han}, \citenamefont {Lee},\ and\
  \citenamefont {Kim}}]{Jun2010}%
  \BibitemOpen
  \bibfield  {author} {\bibinfo {author} {\bibfnamefont {S.~C.}\ \bibnamefont
  {Jun}}, \bibinfo {author} {\bibfnamefont {S.}~\bibnamefont {Moon}}, \bibinfo
  {author} {\bibfnamefont {W.}~\bibnamefont {Kim}}, \bibinfo {author}
  {\bibfnamefont {J.~H.}\ \bibnamefont {Cho}}, \bibinfo {author} {\bibfnamefont
  {J.~Y.}\ \bibnamefont {Kang}}, \bibinfo {author} {\bibfnamefont
  {Y.}~\bibnamefont {Jung}}, \bibinfo {author} {\bibfnamefont {H.}~\bibnamefont
  {Yoon}}, \bibinfo {author} {\bibfnamefont {J.}~\bibnamefont {Shin}}, \bibinfo
  {author} {\bibfnamefont {I.}~\bibnamefont {Song}}, \bibinfo {author}
  {\bibfnamefont {J.}~\bibnamefont {Choi}}, \bibinfo {author} {\bibfnamefont
  {J.~H.}\ \bibnamefont {Choi}}, \bibinfo {author} {\bibfnamefont {M.~J.}\
  \bibnamefont {Bae}}, \bibinfo {author} {\bibfnamefont {I.~T.}\ \bibnamefont
  {Han}}, \bibinfo {author} {\bibfnamefont {S.}~\bibnamefont {Lee}}, \ and\
  \bibinfo {author} {\bibfnamefont {J.~M.}\ \bibnamefont {Kim}},\ }\href
  {\doibase 10.1088/1367-2630/12/4/043023} {\bibfield  {journal} {\bibinfo
  {journal} {N. J. Phys.}\ }\textbf {\bibinfo {volume} {12}},\ \bibinfo {pages}
  {043023--} (\bibinfo {year} {2010})}\BibitemShut {NoStop}%
\bibitem [{\citenamefont {Unterreithmeier}, \citenamefont {Faust},\ and\
  \citenamefont {Kotthaus}(2010{\natexlab{a}})}]{Unterreithmeier2010b}%
  \BibitemOpen
  \bibfield  {author} {\bibinfo {author} {\bibfnamefont {Q.~P.}\ \bibnamefont
  {Unterreithmeier}}, \bibinfo {author} {\bibfnamefont {T.}~\bibnamefont
  {Faust}}, \ and\ \bibinfo {author} {\bibfnamefont {J.~P.}\ \bibnamefont
  {Kotthaus}},\ }\bibfield  {title} {\enquote {\bibinfo {title} {Nonlinear
  switching dynamics in a nanomechanical resonator},}\ }\href {\doibase
  10.1103/PhysRevB.81.241405} {\bibfield  {journal} {\bibinfo  {journal} {Phys.
  Rev. B}\ }\textbf {\bibinfo {volume} {81}},\ \bibinfo {pages} {241405--}
  (\bibinfo {year} {2010}{\natexlab{a}})}\BibitemShut {NoStop}%
\bibitem [{\citenamefont {Timoshenko}(2008)}]{Timoshenko2008}%
  \BibitemOpen
  \bibfield  {author} {\bibinfo {author} {\bibfnamefont {S.}~\bibnamefont
  {Timoshenko}},\ }\href@noop {} {\emph {\bibinfo {title} {Vibration Problems
  In Engeneering}}}\ (\bibinfo  {publisher} {D. Van Nostrand Company, Inc: New
  York},\ \bibinfo {year} {2008})\BibitemShut {NoStop}%
\bibitem [{\citenamefont {Verbridge}\ \emph {et~al.}(2006)\citenamefont
  {Verbridge}, \citenamefont {Parpia}, \citenamefont {Reichenbach},
  \citenamefont {Bellan},\ and\ \citenamefont {Craighead}}]{Verbridge2006}%
  \BibitemOpen
  \bibfield  {author} {\bibinfo {author} {\bibfnamefont {S.~S.}\ \bibnamefont
  {Verbridge}}, \bibinfo {author} {\bibfnamefont {J.~M.}\ \bibnamefont
  {Parpia}}, \bibinfo {author} {\bibfnamefont {R.~B.}\ \bibnamefont
  {Reichenbach}}, \bibinfo {author} {\bibfnamefont {L.~M.}\ \bibnamefont
  {Bellan}}, \ and\ \bibinfo {author} {\bibfnamefont {H.~G.}\ \bibnamefont
  {Craighead}},\ }\href {\doibase 10.1063/1.2204829} {\bibfield  {journal}
  {\bibinfo  {journal} {J. Appl. Phys.}\ }\textbf {\bibinfo {volume} {99}},\
  \bibinfo {pages} {124304--8} (\bibinfo {year} {2006})}\BibitemShut {NoStop}%
\bibitem [{SI()}]{SI}%
  \BibitemOpen
  \href@noop {} {\enquote {\bibinfo {title} {See supplemental material at xxx
  for a rigious derivation of the equations.}}\ }\BibitemShut {NoStop}%
\bibitem [{\citenamefont {Nayfeh}\ and\ \citenamefont
  {Mook}(1979)}]{NonlinearBook2}%
  \BibitemOpen
  \bibfield  {author} {\bibinfo {author} {\bibfnamefont {A.~N.}\ \bibnamefont
  {Nayfeh}}\ and\ \bibinfo {author} {\bibfnamefont {D.}~\bibnamefont {Mook}},\
  }\href@noop {} {\emph {\bibinfo {title} {Nonlinear Oscillations}}}\ (\bibinfo
   {publisher} {John Wiley \& Sons: New York},\ \bibinfo {year}
  {1979})\BibitemShut {NoStop}%
\bibitem [{\citenamefont {Radons}, \citenamefont {Rumpf},\ and\ \citenamefont
  {Schuster}(2010)}]{NonlinearBook}%
  \BibitemOpen
  \bibinfo {editor} {\bibfnamefont {G.}~\bibnamefont {Radons}}, \bibinfo
  {editor} {\bibfnamefont {B.}~\bibnamefont {Rumpf}}, \ and\ \bibinfo {editor}
  {\bibfnamefont {H.~G.}\ \bibnamefont {Schuster}},\ eds.,\ \href@noop {}
  {\emph {\bibinfo {title} {Nonlinear Dynamics of Nanosystems}}}\ (\bibinfo
  {publisher} {WILEY-VCH: Weinheim},\ \bibinfo {year} {2010})\BibitemShut
  {NoStop}%
\bibitem [{\citenamefont {Gorodetsky}\ \emph {et~al.}(2010)\citenamefont
  {Gorodetsky}, \citenamefont {Schliesser}, \citenamefont {Anetsberger},
  \citenamefont {Deleglise},\ and\ \citenamefont
  {Kippenberg}}]{Gorodetsky2010a}%
  \BibitemOpen
  \bibfield  {author} {\bibinfo {author} {\bibfnamefont {M.~L.}\ \bibnamefont
  {Gorodetsky}}, \bibinfo {author} {\bibfnamefont {A.}~\bibnamefont
  {Schliesser}}, \bibinfo {author} {\bibfnamefont {G.}~\bibnamefont
  {Anetsberger}}, \bibinfo {author} {\bibfnamefont {S.}~\bibnamefont
  {Deleglise}}, \ and\ \bibinfo {author} {\bibfnamefont {T.~J.}\ \bibnamefont
  {Kippenberg}},\ }\href {\doibase 10.1364/OE.18.023236} {\bibfield  {journal}
  {\bibinfo  {journal} {Opt. Express}\ }\textbf {\bibinfo {volume} {18}},\
  \bibinfo {pages} {23236--23246} (\bibinfo {year} {2010})}\BibitemShut
  {NoStop}%
\bibitem [{\citenamefont {Unterreithmeier}, \citenamefont {Faust},\ and\
  \citenamefont {Kotthaus}(2010{\natexlab{b}})}]{Unterreithmeier2010a}%
  \BibitemOpen
  \bibfield  {author} {\bibinfo {author} {\bibfnamefont {Q.~P.}\ \bibnamefont
  {Unterreithmeier}}, \bibinfo {author} {\bibfnamefont {T.}~\bibnamefont
  {Faust}}, \ and\ \bibinfo {author} {\bibfnamefont {J.~P.}\ \bibnamefont
  {Kotthaus}},\ }\bibfield  {title} {\enquote {\bibinfo {title} {Damping of
  nanomechanical resonators},}\ }\href {\doibase
  10.1103/PhysRevLett.105.027205} {\bibfield  {journal} {\bibinfo  {journal}
  {Phys. Rev. Lett.}\ }\textbf {\bibinfo {volume} {105}},\ \bibinfo {pages}
  {027205} (\bibinfo {year} {2010}{\natexlab{b}})}\BibitemShut {NoStop}%
\bibitem [{\citenamefont {Cleland}(2003)}]{cleland_foundations_2003}%
  \BibitemOpen
  \bibfield  {author} {\bibinfo {author} {\bibfnamefont {A.~N.}\ \bibnamefont
  {Cleland}},\ }\href@noop {} {\emph {\bibinfo {title} {Foundations of
  Nanomechanics}}}\ (\bibinfo  {publisher} {Springer},\ \bibinfo {year}
  {2003})\BibitemShut {NoStop}%
\end{thebibliography}


\end{document}